\documentstyle[12pt]{article}
\oddsidemargin = - 6 pt
\begin{document}
\begin{center}
{\large \bf  On B\"{a}cklund transformations and boundary conditions associated with the quantum inverse problem for a discrete nonlinear integrable system and its connection to Baxter's $Q$-operator}
\end{center}
\vskip 50 pt
\begin{center}{ \bf A. Ghose Choudhury,\\}
{\it Department of Physics, Surendranath College,\\}
{\it 24/2 Mahatma Gandhi Road, Calcutta-700 009, India\\}
{ \it email: aghose@cal3.vsnl.net.in\\}
 and\\
{\bf  A. Roy Chowdhury  \\}
{ \it High Energy Physics Division  \\
Department of Physics, Jadavpur University  \\ 
Calcutta -700 032, India.\\}
{\it  email: arcphy@cal2.vsnl.net.in\\}
\end{center}
\vskip 10 pt
\begin{center}
{\bf Abstract}
\end{center}
A discrete nonlinear system is analysed in case of open chain boundary conditions at the ends. It is shown that the integrability of the system remains intact, by obtaining a modified set of Lax equations which automatically take care of the boundary conditions. The same Lax pair also conforms to the conditions stipulated  by Sklyanin [5]. The quantum inverse problem is set up and the diagonalisation is carried out by the method of sparation of variables. B\"{a}cklund transformations are then derived under the modified boundary conditions using the classical $r$-matrix . Finally by quantising the B\"{a}cklund transformation it is possible to identify the relation satisfied by the eigenvalue of Baxter's $Q$-operator even for the quasi periodic situation.
\newpage
\section{Introduction}
Analysis of nonlinear integrable systems is a subject of immense importance both from the physical and mathematical points of view. The inverse problem formulated on the basis of Lax operators can solve a wide class of  problems. On the other hand, it is also important that  problems  be formulated with pre-assigned boundary conditions and solutions of the corresponding Cauchy problem be obtained.  However at times the imposition of  finite boundary conditions may lead to a loss of integrability. An important developement in this regard was the discovery by Sklyanin [1], who showed how non-periodic boundary conditions could be imposed on an integrable model without destroying its integrability. In this communication we have analysed a discrete nonlinear system, (also known as the DST model), initially studied by Ragnisco {\it et al} and also recently by Sklyanin {\it et al} from the view point of its integrability in presence of open boundary conditions. Our approach is different from that of Sklyanin in the sense that we have changed the form of the Lax operator at the end points where the boundary conditions are imposed. We have then shown that our approach can also accomodate the conditions laid down by Sklyanin for the system to be integrable under such circumstances. The present formalism was first used by Zhou [3] for a fermionic spin system. Both the classical and quantum hamiltonians have been derived explicitly. The quantum $R$ matrix has also been deduced so as to formulate the quantum inverse problem for the model under finite boundary conditions. Diagonalisation of the hamiltonian is  carried out by using the method of separation of variables,  as the boundary matrices are non diagonal in character. Of late there has been a great deal of interest in the study of canonical B\'{a}cklund transformations within the framework of classical $r$ matrix theory, specially under periodic boundary conditions [15]. We have discussed this issue in case of quasi -periodic and finite boundary conditions. Finally in the case of the less stringent quasi-periodic boundary conditions we have shown how a connection with the eigenvalue of Baxter's $Q$ operator may be established [11] by quantising the B\"{a}cklund transformation.\\
The model under analysis is described by the following equations of motion.
$$ \dot q_n =q_{n+1} -q_n^2 r_n$$
$$ \dot r_n = -r_{n-1} +q_n r_n^2 \eqno(1.1)$$ where $ \dot q_n $ stands for the time derivative of $q_n$.  It was originally proposed by Ragnisco {\it et al} [2] in their analysis of Lie-B\"{a}cklund symmetries of discrete systems. The Lax pair associated with (1.1) may be written as 
$$\Psi_{n+1}=L_n\Psi_n  \hskip 30pt \Psi _{n t}= M_n \Psi _n\eqno(1.2)$$ where 
$$ L_n(\lambda )=\left(\begin{array}{cccc}
\lambda +q_n r_n & q_n\\ r_n & 1\end{array}\right)\hskip 20pt
 M_n(\lambda )=\left(\begin{array}{cccc}
\frac{\lambda }{2} & q_n\\ r_{n-1} & -\frac{\lambda }{2}\end{array}\right)\eqno(1.3)$$
Note that consistency of (1.2) yields the equations of motion only for periodic boundary conditions. If however certain nontrivial boundary conditions are to be introduced then one has to adopt a different strategy as will be explained in the sequel. The system given by (1.1) possesses a hamiltonian structure with the following symplectic form [4].
$$\left(\begin{array}{ccc}
q_n\\ r_n\end{array}\right)_t =\left(\begin{array}{cccc}
0 & 1\\ -1 & 0\end{array}\right) \left(\begin{array}{cccc}
\frac{\delta  H}{\delta q_n}\\ \frac{\delta H}{\delta r_n}\end{array}\right)\hskip 10 pt
H=\frac{1}{2}\sum_n (q_{n+1} r_n + q_n r_{n-1} -q_n^2 r_n^2)\eqno(1.4)$$
For the purpose of introducing the boundary conditions we modify the Lax operator in (1.2) in the following manner,
$$ \Psi _{j+1} =L_j (\lambda )\Psi _j, \hskip 10 pt j=1,2,....N$$
$$\frac{d\Psi _j}{dt} =M_j(\lambda )\Psi _j \hskip 30 pt j=2,3,....N-1$$
$$\frac{d\Psi _{N+1}}{dt}  =W_{N+1}(\lambda ) \Psi _{N+1}$$
$$ \frac{d\Psi _1}{dt}=W_1(\lambda )\Psi _1\eqno(1.5)$$ where $W_{N+1}, W_1$ are two new $2\times2$ matrices depending on the spectral parameter $\lambda $ and on the dynamical variables.
The usual consistency condition of (1.2) viz
$$ \frac{dL_j(\lambda )}{dt} =M_{j+1} L_j(\lambda ) -L_j(\lambda ) M_j(\lambda )\eqno( 1.6)$$ is now replaced by the following set of equations:
$$\frac{ dL_j(\lambda )}{dt} =M_{j+1}(\lambda ) L_j(\lambda ) -L_j(\lambda ) M_j(\lambda )\hskip 10pt (j=1,2....N-1)$$
$$\frac{dL_N(\lambda )}{dt} =W_{N+1}L_N(\lambda ) -L_N(\lambda )M_N(\lambda )$$
$$\frac{dL_1(\lambda )}{dt} =M_2(\lambda ) L_1(\lambda ) -L_1(\lambda ) W_1(\lambda )\eqno(1.7)$$
That is the consistency conditions at the two ends are different. In the present case we have formally.
$$ W_{N+1}=\left(\begin{array}{cccc}
\frac{\lambda }{2} & q_{N+1}\\
r_N & -\frac{\lambda }{2}\end{array}\right) \hskip 30pt  W_1=\left(\begin{array}{cccc}
\frac{\lambda }{2} & q_1\\
r_0 & -\frac{\lambda }{2}\end{array}\right) \eqno(1.8)$$ so that upon imposing the boundary conditions $r_0=\theta _{-}, q_{N+1}=\theta _{+}$ the equations for the two ends of the discrete chain turn out to be as follows.
$$\dot q_1= q_2 -q_1^2 r_1\hskip 20pt \dot r_1=-(\theta _{-} -q_1 r_1^2)$$
$$ \dot q_N =(\theta _{+} -q_N^2 r_N)\hskip 20pt \dot r_N=-(r_{N-1}-q_N r_N^2)\eqno(1.9)$$ 
We shall refer to the above system as an open chain.
\section{ Boundary conditions and classical $r-$ matrix}
To ascertain the hamiltonian associated with the open chain system it is always useful to compute the so called {\it classical r-matrix}  through the poisson  bracket given by the symplectic form in (1.4) [8]. It is straight forward to show that.
$$\{ L_n(\lambda )\otimes_, L_m(\mu )\}=[r(\lambda , \mu ), L_n(\lambda )\otimes L_n(\mu )]\delta_{nm}\eqno(2.1)$$ where 
$ r(\lambda , \mu ) =-\frac{{\cal P}}{\lambda -\mu }, {\cal P}$ standing for the permutation operator. The {\it mondromy} matrix is then defined by
$$T_N(\lambda )=\prod_{n=1}^N L_n(\lambda )\eqno( 2.2)$$ From (2.1) and (2.2) it follows that
$$\{ T_N(\lambda )\otimes_, T_N(\mu )\}=[r(\lambda , \mu ), T_N(\lambda )\otimes T_N(\mu )]\eqno(2.3)$$ We shall refer to this poisson algebra as the classical inverse scattering method (CISM-I) algebra. It follows that
$$\frac{dT_N}{dt}(\lambda ) =\sum_{n=1}^N L_N...L_{n+1} \dot L_n L_{n-1}....L_1 (\lambda ) =W_{N+1} T_N(\lambda ) -T_N(\lambda ) W_1\eqno(2.4)$$ from which it is easy to conclude that $J(\lambda )=tr T_N(\lambda ) $ is a constant of motion in the periodic case {\it i.e} when $q_{m+N}=q_m$ and $r_{m+N}=r_m$.
\subsection{Quasiperiodic case}
Instead of the periodic boundary conditions that are customarily assumed, one can impose the so called {\it quasiperiodic boundary conditions} on the system, without destroying its integrability. This requires the existence of a matrix $C(\lambda , \xi )$ such that
$$\bar J(\lambda ) =tr [C(\lambda ,\xi ) T_N(\lambda )]\eqno(2.5)$$ is time independent so that in the present situation one has
$$ \frac{d\bar J(\lambda )}{dt}=tr [C(\lambda , \xi ) (M_{N+1} T_N- T_N M_1)]=tr[C(\lambda ,\xi ) M_{N+1}T_N] - tr [M_1 C(\lambda, \xi)T_N]\eqno(2.6)$$ from which it follows that  
$$ C(\lambda , \xi ) M_{N+1} = M_1 C(\lambda , \xi )\eqno(2.7)$$ As a specific example one can take 
$$C(\xi )=\left(\begin{array}{cccc}
\xi ^{-\frac{1}{2}} & 0\\
0 &  \xi ^{\frac{1}{2}}\end{array}\right)\eqno(2.8)$$ To analyse the structure of $\bar J(\lambda )$ in detail, we note the following asymptotics of the monodromy matrix in our case.
$$ T_N^{11}(\lambda ) \approx \lambda ^N +\lambda ^{N-1} S + p_2 \lambda ^{N-2}+....$$
$$T_N^{12}(\lambda )\approx\lambda ^{N-1} q_1 +O(\lambda ^{N-2})$$
$$T_N^{21}(\lambda )\approx \lambda ^{N-1} r_N +O(\lambda ^{N-2})$$
$$ T_N^{22}\approx \lambda ^{N-2} p_2^{\prime} +O(\lambda ^{N-2})\eqno(2.9)$$ whence we get 
$$\bar J(\lambda )=tr (C(\xi )T_N(\lambda))=\xi ^{-\frac{1}{2}}\lambda ^N +(\xi ^{-\frac{1}{2}}S)\lambda ^{N-1} + (\xi ^{-\frac{1}{2}}p_2)\lambda ^{N-2} +
(\xi ^{\frac{1}{2}}p_2^{\prime})\lambda ^{N-2} +....\eqno(2.10)$$ where 
$$ S=\sum_{i=1}^N s_i , \hskip 10pt p_2 =\sum_{i=1}^{N-1} (q_{i+1} r_i) +\sum_{i<j} s_i s_j, \hskip 10pt p_2^{\prime}= r_N q_1 \eqno( 2.11)$$One can verify, using the equations of motion, that each coefficient of $\lambda ^j$ is a constant. The following nonlinear combination of these constants then gives the Hamiltonian in the quasiperiodic case.
$$ H=-\frac{1}{2}\xi ^{-\frac{1}{2}}(S^ 2 -2p_2) +\xi ^{\frac{1}{2}} p_2^{\prime} =
\xi ^{-\frac{1}{2}}[\sum_{i=1}^{N-1} q_{i+1} r_i -\frac{1}{2} \sum_i q_i^2 r_i^2 +\xi r_N q_1]\eqno(2.12)$$
\subsection{Classical open chain}
As mentioned earlier, the open chain in which independent boundary conditions are imposed at the two ends, represents a more general case compared to the quasi-periodic boundary conditions. The equations of motion are given by ( 1.1) and ( 1.9). Now by following the arguments given by Sklyanin in [1] one can show that 
$$\tau (\lambda ) =tr [K_{+}(\lambda ) U(\lambda )], \hskip 20 pt U(\lambda )=T_N(\lambda )K_(\lambda )T_N^{-1}(-\lambda )\eqno( 2.13)$$ represents the generator of the infinite conserved quantities with the matrices $K_{\pm}(\lambda )$ incorporating the effects of the nontrivial boundary conditions. In our case the classical $r-$ matrix being symmetric, both $K_{\pm}$ satisfy the same equation namely
$$[r(\lambda , \mu ), K(\lambda )\otimes  K(\mu )]+K^1(\lambda )r(\lambda +\mu )K^2(\mu ) -K^2(\mu ) r(\lambda +\mu )K^1(\mu )=0\eqno(2.14)$$ where $K^1(\lambda )=K(\lambda )\otimes  I$ and $K^2(\lambda )=I\otimes K(\lambda )$. The following solutions of (2.14) are  found to be admissible .
$$K_{-}(\lambda )=\left(\begin{array}{cccc}
\theta _{-} & \lambda \\ 0 & \theta _{-}\end{array}\right) \hskip 30 pt K_{+}(\lambda ) =\left(\begin{array}{cccc}
\theta _{+} & 0\\ \lambda  & \theta _{+}\end{array}\right)\eqno( 2.15)$$ By explicit multiplication we find using (2.9) and ( 2.15) in ( 2.13) 
$$\tau (\lambda ) =\lambda ^{N+2} + 2\lambda ^N ( a_2 -\frac{1}{2} S^2 +q_1 \theta _{-} + r_N\theta _{+}) +O(\lambda ^{N-2})\eqno( 2.16)$$ where $ a_2 =\sum_{i<j=1}^{N} s_i s_j +\sum _{i=1}^{N-1} q_{i=1} r_i$ and $S=\sum_{i=1}^N s_i$ so that  the hamiltonian is given by the  expression.
$$ H=\sum_{i=1}^{N-1} q_{i+1} r_i -\frac{1}{2} \sum_{i=1}^N q_i^2 r_i^2 +q_1\theta _{-} + r_N \theta _{+}\eqno(2.17)$$ To ascertain  if $\dot \tau (\lambda )=0$ we observe that 
$$\frac{d\tau }{dt}(\lambda ) =tr [K_{+}(\lambda )\dot T_N(\lambda ) K_{-}(\lambda )T_N^{-1}(-\lambda )+ K_{+}(\lambda )T_N(\lambda )K_{-}(\lambda )(-T_N^{-1}(-\lambda )\dot T_N(-\lambda )T_N^{-1}(-\lambda ))]$$
$$ \frac{d\tau }{dt}(\lambda ) =tr [K_{+}(\lambda ) W_N(\lambda )U(\lambda )] - tr[K_{+}(\lambda ) T_N(\lambda )W_1(\lambda )K_{-}(\lambda )T_N^{-1}(-\lambda )]$$
$$-tr[K_{+}(\lambda )U(\lambda )W_N(-\lambda )] +tr[K_{+}(\lambda ) T_N(\lambda )K_{-}(\lambda )W_1(-\lambda )T_N^{-1}(-\lambda )]\eqno(2.18)$$ hence  $\dot \tau (\lambda )=0$ if and only if
$$ K_{+}(\lambda )W_{N+1}(\lambda )=W_{N+1}(-\lambda )K_{+}(\lambda )\eqno(2.19)$$
$$ W_1(\lambda )K_{-}(\lambda )=K_{-}(\lambda )W_1(-\lambda )\eqno(2.20)$$ which are found to be identical to the conditions deduced by Sklyanin [5].
\section{B\"{a}cklund transformation}
In this section we derive a B\"{a}cklund transformation (BT) for the model under consideration, using the recently developed approach of Sklyanin [6, 7]. This relies on the algebra of the local lax operator as given by (2.1). For the sake of clarity we shall begin by retracing the methods used by Sklyanin in [16] and then consider the case where the nonlinear variables do not obey periodic boundary conditions. Sklyanin's formalism requires us to look for another representation of the local algebra given in (2.1). Let $g_i(\lambda -\sigma )$ be such a representation satisfying
$$\{ g_i(\lambda -\sigma )\otimes_, g_i(\mu -\sigma )\} =[r(\lambda -\mu ), g_i(\lambda -\sigma )\otimes g_i(\mu -\sigma )]\eqno(3.1)$$where $\lambda $ and $\mu $ are spectral parameters and $\sigma $ is a complex parameter of the transformation. Moreover $g_i(\lambda -\sigma )$ is assumed to depend on a canonical set of variables  $(t_i, T_i)$. In (3.1)  $r(\lambda -\mu )=-\frac{{\cal P}}{\lambda -\mu }$. To construct the BT 
$$B_{\sigma } : T_N(x, X ;\lambda )\longrightarrow T_N(y, Y; \lambda )\eqno(3.2)$$ we shall first construct a local transformation
$$b_{\sigma }^{(i)}:L_i(x_i, X_i; \lambda )\longrightarrow L_i( y_i, Y_i; \lambda )\eqno(3.3)$$ defined by the following
$$g_i(\lambda -\sigma ; s_i , S_i)L_i(\lambda ; x_i, X_i) =L_i(\lambda ; y_i, Y_i) g_i(\lambda -\sigma ; t_i, T_i)\eqno(3.4)$$ here $L_i(\lambda ; x_i, X_i)$ is the same as in (1.3), we have simply relabelled the variables $q_i, r_i$ by $x_i, X_i$ for notational convenience. In (3.4) we see that in addition to the canonical variables $(x_i, X_i)_{i=1}^N$ and $(y_i, Y_i)_{i=1}^N$ defining the BT we have two additional sets of canonical variables 
$(s_i, S_i)_{i=1}^N$  and $(t_i, T_i)_{i=1}^N$. It is assumed that all the four sets of canonical variables are mutually commuting quantities. The transformation $b_{\sigma }^{(i)}$ is therefore a transformation in an extended phase space. Now by  imposing suitable constraints it is possible to eliminate the auxillary variables $(s_i, S-i)$ and $(t_i, T_i)$ from (3.4) thereby obtaining a local transformation $b_{\sigma }^{(i)}$. We get from (3.4) the following set of independent equations, taking $g_i(\lambda -\sigma ; s_i, S_i)$ to be of the form.
$$ g_i(\lambda -\sigma ); s_i, S_i) =\left(\begin{array}{cccc} 1 & s_i\\ -S_i & \lambda -\sigma -s_i S_i\end{array}\right)\eqno(3.5)$$ 
$$(x_i +s_i) X_i=y_i(Y_i-T_i), \hskip 20pt -S_iX_i(x_i+s_i)=Y_i-T_i +\sigma X_iy_i$$
$$t_i=-y_i, \hskip 30pt X_i=S_i\eqno(3.6)$$Imposition of the constraints $s_i=t_{i+1}$ and $ S_i=T_{i+1}$ in these equations leads to 
$$X_i=-\frac{1}{y_i} -\frac{\sigma }{x_i-y_{i+1}}\hskip 20pt Y_i=X_{i-1}+\frac{x_i-y_{i+1}}{y_i}X_i\eqno(3.7)$$ which defines the local transformation $b_{\sigma }^{(i)}:(x_i, X_i)\rightarrow(y_i, Y_i)$.  It should be pointed out that this local BT may be obtained from a generating function $G_\sigma $ where
$$G_{\sigma }=\sum_{i=1}^N \ [\frac{x_i-y_{i+1}}{y_i} +\sigma \log (\frac{x_i-y_{i+1}}{y_i})\ ]\eqno(3.8)$$ $$ X_i=-\frac{\partial G_\sigma }{\partial x_i},\hskip 30pt Y_i=\frac{\partial G_\sigma }{\partial y_i}\eqno(3.9)$$ Moreover as a consequence of the constraints $s_i=t_{i+1}, S_i=T_{i+1}$ and (3.6) we see that 
$$g_i(\lambda -\sigma ; s_i, S_i)=g_{i+1}(\lambda -\sigma ; t_i, T_i)\eqno(3.10)$$ and hence (3.4) becomes
$$ g_{i+1}(\lambda -\sigma ; y_{i+1},X_i) L_i(\lambda ; x_i, X_i) =L_i(\lambda ; x_i X_i) g_i(\lambda -\sigma ; y_i, X_{i-1})\eqno(3.11)$$ Now it is well known that the generating function for the integrals of motion are given by the transfer  matrix $J(\lambda )$ defined as the trace of the mondromy matrix, {\it i.e}
$$J(\lambda ) =tr T_N(x, X;\lambda )\eqno(3.12)$$ where $T_N(\lambda )$ obeys the CISM-I algebra.
$$ \{T_N(\lambda )\otimes_, T_N(\mu )\}=[r(\lambda -\mu ), T_N(\lambda )\otimes T_N(\mu )]\eqno(3.13)$$ Under the BT defined by (3.11) the transfer matrix assumes the form.
$$J(\lambda )=tr[g_{N+1}^{-1} (\lambda -\sigma ; y_{N+1},X_N)T_N(y,Y;\lambda )g_1(\lambda -\sigma ;y_1,X_0)]\eqno(3.14)$$
Clearly invariance of the integrals of motion and thus of the hamiltonian generated by $J(\lambda )$ requires that the variables in $g_i(\lambda -\sigma )$ be periodic {\it i.e} $$X_{N+m}=X_m, y_{N+m}=y_m\eqno(3.15)$$ In such cases the BT will be canonical in nature since the hamiltonian is invariant.\\
In the quasi periodic case the boundary conditions on the dynamical variables are determined by (2.7), so that upon using the form of $C(\xi )$ as given in (2.8) we find that the variables in $L_i(\lambda )$ obey
$$x_{N+1}=\xi X_N, \hskip 30pt X_0=\xi X_{N} \eqno( 3.16)$$ We refer to these as the quasi periodic boundary conditions. Now in view of (2.5) which defines the generator of the integrals of  motion $\bar J(\lambda )$ we see that as a consequence of the BT given by (3.12), it is modified to the following.
$$\bar J(\lambda )=tr [C(\lambda ,\xi ) g_{N+1}^{-1}(\lambda -\sigma ; y_{N+1}, X_N)T_N(y, Y;\lambda ) g_1(\lambda -\sigma ; y_1, X_0)]\eqno( 3.17)$$ Clearly invariance of $\bar J(\lambda )$ requires that 
$$ g_1(\lambda -\sigma ; y_1, X_0)C(\lambda ,\xi ) g_{N+1}^{-1}(\lambda -\sigma ; y_{N+1}, X_N)=C(\lambda ,\xi )\eqno(3.18)$$ which gives.
$$y_{N+1}=\xi y_1,\hskip 30pt X_0=\xi X_N\eqno(3.19)$$ Structurally these equations are consistent with (3.16) because the Lax operator $L_i(\lambda ; y_i, Y_i)$ in (3.11) is obtained simply by replacing the variables $(x_i, X_i)_{i=1}^N$ with $(y_i, Y_i)_{i=1}^N$ .  Therefore the BT generated by (3.11) through the matrix $g_I(\lambda -\sigma )$ is a canonical transformation even when the nonlinear dynamical variables obey quasiperiodic boundary conditions. It is now natural to enquire if a BT can be constructed in case of open chains. Infact the most general type of boundary conditions for  Liouville integrable lattices was formulated some years back by Sklyanin in terms of the representations of a new algebra known as the reflection equation algebra (REA). The classical limit of the REA is given by the following Poisson algebra, which shall be referred to as the CISM-II algebra.
$$\{T^{(1)}(\lambda ), T^{(2)}(\mu )\}=[r(\lambda -\mu ), T^{(1)}(\lambda )T^{(2)}(\mu )] +T^{(1)}(\lambda )r(\lambda +\mu )T^{(2)}(\mu ) -T^{(2)}(\mu )r(\lambda +\mu )T^{(1)}(\lambda )\eqno(3.20)$$Now if $K_{\pm}(\lambda )$ be representations of the above algebra then it may be shown that $U(\lambda )=T_N(\lambda )K_{-}(\lambda )T_N^{-1}(-\lambda )$ also satisfies the CISM-II algebra, where $T_N(\lambda )$ is defined by (2.2). Consequently it follows that 
$$\tau (\lambda )=tr K_{+}(\lambda )U(\lambda )\eqno( 3.21)$$ is a generating function of the integrals of motion. We have already seen that under the BT given  by (3.11) the monodromy matrix undergoes the following transformation.
$$B_{\sigma }: T_N(x, X;\lambda )\longrightarrow g_{N+1}^{-1}(\lambda -\sigma ; y_{N+1}, X_N)T_N(y, Y; \lambda )g_1(\lambda -\sigma ; y_1, X_0)\eqno(3.22)$$ Next consider $2\times 2$ matrices $V_{\pm}(\lambda )$ and construct the following quantities:
$$Z(\lambda )\equiv T_N(x, X;\lambda )V_{-}(\lambda )T_N^{-1}(x, X;-\lambda )\hskip 20pt \tilde J(\lambda )\equiv tr (V_{+}(\lambda )Z(\lambda ))\eqno(3.23)$$ note that $Z(\lambda )$ per se might not satisfy the CISM -II algebra. But if we apply a BT as given by (3.11) then $\tilde J(\lambda )$ assumes the following form.
$$ \tilde J(\lambda)=tr [\{g_{N+1}(-\lambda -\sigma )V_{+}(\lambda)g_{N+1}^{-1}(\lambda -\sigma )\} T_N(y, Y; \lambda)\times$$
$$\times\{g_1(\lambda-\sigma)V_{-}(\lambda)g_1^{-1}(-\lambda -\sigma )\}T_N^{-1}(y, Y;-\lambda )]\eqno(3.24)$$ It is evident that $\tilde J(\lambda )$ would be a generator of the integrals of motion provided that the quantities in curly brackets are representations of the CISM -II algebra. This infact opens up the possibility of deriving a hierarchy of suitable boundary conditions for integrable systems starting from appropriate matrices $V_{+}$. We will examine this possibility by deriving the usual scalar boundary matrices. First of all if the quantities in curly brackets were similarity transformations, then we could have taken the matrices $V_{\pm}$ to be simply the matrices $K_{\pm}(\lambda )$ as defined in ( 2.15). This is a consequence of the observation that the CISM-II algebra is invariant under a similarity transformation of its elements. However this possibility is spoilt by the fact that the arguments of $g_{1(N+1)}$ and $g_{1(N+1)}^{-1}$ in ( 3.24) are different. Consider therefore the following forms of the matrices $V_{\pm}(\lambda )$ 
$$ V_{+}(\lambda ) =-\frac{\lambda }{\lambda +\sigma }\left(\begin{array}{cccc}
a+\frac{d}{\lambda } & b\\ 1 & -a +\frac{d}{\lambda }\end{array}\right)\eqno( 3.25)$$ 
$$ V_{-}(\lambda ) =-\frac{\lambda }{\lambda -\sigma }\left(\begin{array}{cccc}
\lambda  A_1 +A_0+\frac{\delta}{\lambda } & \beta \lambda ^2 +B_0\\ C_0 & \lambda  A_1 -A_0 +\frac{\delta }{\lambda }\end{array}\right)\eqno( 3.26)$$ so that 
$$ g_{N+1}(-\lambda -\sigma ) V_{+}(\lambda ) g_{N+1}^{-1}(\lambda -\sigma )=K_{+}(\lambda )=\left(\begin{array}{cccc} \theta _{+} & 0\\ \lambda & \theta _{+}\end{array}\right)\eqno( 3.27)$$ and 
$$ g_1(\lambda -\sigma )V_{-}(\lambda ) g_1^{-1}(-\lambda -\sigma )=K_{+}(\lambda )=\left(\begin{array}{cccc}
\theta _{-} & \lambda \\ 0 & \theta _{-}\end{array}\right)\eqno( 3.28)$$  Here
$$ a=y_{N+1}-\theta_{+}\hskip 20pt d=-\sigma \theta _{+}, \hskip 20pt b=y_{N+1}( 2\theta _{+} -y_{N+1})\eqno( 3.29)$$ and \hskip 20pt $ A_1= X_0, \hskip 10pt A_0=-(\sigma X_0 +\theta_{-}-y_1 X_0^2), \hskip 20pt \delta =\sigma \theta_{-}$ 
$$ \beta =1, \hskip 10pt C_0=X_0^2,\hskip 10pt B_0=-(y_1X_0 -\sigma)^2 +2\theta_{-} y_1\eqno( 3.30)$$ Disregarding the overall factor in $V_{+}(\lambda)$ we see that its structure is similar to the general scalar solution of the CISM -II algebra. While that of $ V_{-}(\lambda)$ is similar to that derived in [10] ( compare their eqns ( 2.3) -(2.5) with $\alpha =\gamma =0, \beta =1)$. The difference in our case is that all the entries in $ V_{-}(\lambda )$ are scalars. Thus while the former is by itself a solution of the CISM -II algebra, for $V_{-}(\lambda)$ to do the same it is necessary that $X_0=\theta_{-}=0$. The other pertinent issue is whether the matrices $V_{\pm}(\lambda)$ are compatible with the integrability of the model under consideration. the constraint of integrability requires these matrices to satisfy (2.20) which is a stringent condition. In case of $V_{+}(\lambda)$ the boundary condition $y_{N+1}=\theta_{+}$ ensures $a=b=0$ so that the integrability condition is not fulfiled unless $Y_N=constant$ which clearly is in contradiction to the fact that $(y_i, Y_i)_{i=1}^N$ are dynamical variables. Thus $V_{\pm}(\lambda)$ by themselves cannot be made to satisfy the requirements of integrabilty and CISM -II algebra simultaneously. However as a result of the BT the quantities $Z(\lambda)$ and $\tilde J(\lambda)$ are modified in a such a manner that the new boundary matrices $ K_{\pm}(\lambda)$ appearing in lieu of $V_{\pm}(\lambda)$ are not only solutions of CISM -II algebra but are also compatible with the integrability of the model under consideration.
\section{Quantum open chain}
In this section we consider the quantum inverse problem for a open chain. Note that the Lax operator in 
(1.3) is essentially similar to the discrete self trapping (DST) model [16]. The integrability and quantum inverse problem in absence of finite boundary conditions were studied in [12, 13]. From the hamiltonian structure of our system we have seen that $( q_n, r_n)$ are a conjugate pair of variables; so in the quantum version of the model we impose the following commutation relations.
$$ [q_n, r_m]=\eta \delta _{nm}\hskip 10 pt (\eta =i\hbar )\hskip 20 pt [q_n, q_m]=[r_n, r_m]=0\eqno(4.1)$$ As before the monodromy matrix is defined by 
$ T_N(\lambda )=\prod_{n=1}^N L_n(\lambda )$ and obeys the QISM -I algebra [8]
$$R_{12}(\lambda -\mu )T_N^{1}(\lambda ) T_N^{2}(\mu ) =T_N^{2}(\mu )T_N^{1}(\lambda )R_{12}(\lambda -\mu )\eqno( 4.2)$$ with $ R_{12}(\lambda -\mu )={\cal I} +\frac{\eta }{\lambda -\mu }{\cal P}$. The matrices determining the boundary conditions $K_{\pm }(\lambda )$ obey the conditions [9].
$$ R_{12}(\lambda -\mu ) K_{-}^{(1)}(\lambda )R_{12}(\lambda +\mu ) K_{-}^{(2)}(\lambda )=K_{-}^{(2)}(\mu )R_{12}(\lambda +\mu )K_{-}^{(1)}R_{12}(\lambda -\mu )\eqno(4.3)$$
$$ R_{12}(-\lambda +\mu )K_{+}^{(1)t_1}(\lambda ) R_{12}(-\lambda -\mu -2\eta ) K_{+}^{(2)t_2}(\mu)$$
$$=K_{+}^{(2)t_2}(\mu)R_{12}(-\lambda -\mu -2\eta ) K_{+}^{(1)t_1}(\lambda ) R_{12}(-\lambda +\mu )\eqno( 4.4)$$ The form of $K_{\pm }(\lambda )$ given in (2.15) is seen to satisfy these equations. The generators of the conserved densities are then built from.
$$ U(\lambda )=\left(\begin{array}{cccc} A(\lambda ) & B(\lambda )\\
C(\lambda ) & D(\lambda )\end{array}\right)=T(\lambda )K_{-}(\lambda -\eta /2, \xi _{-}) \sigma _2 T^{t}(-\lambda )\sigma _2\eqno( 4.5)$$ which obeys the algebra
$$R_{12}(\lambda -\mu ) U^{1}(\lambda )R_{12}(\lambda +\mu -\eta )U^{2}(\lambda )=U^{2}(\lambda )R_{12}(\lambda +\mu -\eta )U^{1}(\lambda )R_{12}(\lambda -\mu )\eqno(4.6)$$ and is given by
$$\tau (\lambda )=tr [K_{+}(\lambda -\eta /2, \xi _{+})U(\lambda )]=\lambda ^{2N+2} +2\lambda^{2N}\hat H_q+..........\eqno( 4.7)$$ where the hamiltonian with quantum corrections is given by 
$$\hat H_q=\sum_{i=1}^{N-1} q_{i+1}r_i -\frac{1}{2}\sum_{i=1}^N(q_ir_i)^2 -\frac{\eta ^2}{8} +\xi _{+}r_N +\xi _{-}q_1\eqno( 4.8)$$ On the other hand with $U(\lambda )$ given by (4.5), we can write the general form of $\tau (\lambda )$ as [10] $$\tau (\lambda )=\xi _{+}(A(\lambda ) +D(\lambda )) +(\lambda +\eta /2)B(\lambda )\eqno( 4.9)$$
\subsection{Separation of variables} From (4.6) we can find the operator algebra for the elements $A(\lambda ), B(\lambda )$ etc. It is however more convenient to define instead of $D(\lambda )$ the following operator.
$$D^{*}(\lambda )=2\lambda D(\lambda )-\eta A(\lambda )\eqno(4.10)$$ The commutation relations relevant for our purposes are then obtainable from (4.3) and are as follows.
$$ [B(\lambda ), B(\mu )]=0\eqno( 4.11)$$
$$A(\lambda )B(\mu ) = \frac{(\lambda -\mu -\eta )(\lambda +\mu -\eta) }{(\lambda -\mu )(\lambda +\mu) }B(\mu )A(\lambda ) +\frac{\eta (2\mu -\eta )}{(\lambda -\mu )2\mu }B(\lambda )A(\mu ) - \frac{\eta }{(\lambda +\mu )2\mu}B(\lambda )D^{*}(\mu )\eqno(4.12)$$
$$ D^{*}(\lambda )B(\mu )= \frac{(\lambda -\mu +\eta )(\lambda +\mu -\eta )}{(\lambda -\mu )(\lambda +\mu )} B(\mu )D^{*}(\lambda ) +$$
$$+\frac{\eta (2\lambda +\eta )(2\mu -\eta )}{(\lambda +\mu )2\mu} B(\lambda )A(\mu ) -\frac{\eta (2\lambda +\eta)}{(\lambda-\mu)2\mu} B(\lambda )D^{*}(\mu )\eqno( 4.13)$$ Furthermore from the definition of $U(\lambda)$ as given in (4.5) the operators $A(\lambda), B(\lambda)$ etc are found to have the following forms:
$$ A(\lambda)=(-)^N r_N\{ \lambda^{2N} +(S+\eta/2)\lambda^{2N-1}+........\}\eqno( 4.14)$$ 
$$ D(\lambda)=(-)^N r_N\{ \lambda^{2N} +(S-\eta/2)\lambda^{2N-1}+........\}\eqno( 4.15)$$ 
$$B(\lambda) =(-)^N (\lambda -\eta/2)\{ \lambda^{2N} +......\}\equiv (-)^N (\lambda -\eta/2)\prod_{\alpha =1}^N (\lambda -\hat u_\alpha )(\lambda +\hat u_{\alpha })\eqno( 4.16)$$  Where $S=\sum_{i=1}^N q_i r_i$. From ( 4.9) we note that the presence of the operator $B(\lambda)$ in the expression for the transfer matrix $\tau(\lambda)$ prevents us from setting up the usual algebraic Bethe ansatz. Consequently we have recast $B(\lambda)$ in the  form given in (4.16) where $\hat u_{\alpha}$ are the zeros of this operator. Henceforth $\{ \hat u_\alpha \mid \alpha=1, 2...N\}$ are themselves to be regarded as operators defining $B(\lambda)$ and it follows from their commutativity given by (4.11) that 
$$ [\hat u_\alpha , \hat u_\beta ]=0\eqno( 4.17)$$ We next define operators $\hat v_{\alpha }^{\pm }$ by left substitution $\lambda \Rightarrow \hat u_\alpha $ into the operators $A(\lambda )$ and $D^{*}(\lambda)$ as shown below :
$$\hat v_\alpha ^{+}\equiv D^{*}(\lambda \Rightarrow \hat u_\alpha ), \hskip20 pt \hat v_\alpha ^{-}\equiv A(\lambda \Rightarrow \hat u_\alpha )\eqno( 4.18)$$ It then follows from the algebra of the operators $A(\lambda ), D^{*}(\lambda )$ that 
$$[\hat v_\alpha ^{\pm }, \hat u_\beta ]=\pm \eta \hat v_\alpha ^{\pm }\delta _{\alpha \beta } \hskip 10pt [\hat v_{\alpha }^{+}, \hat v_\beta^{+}]=0 \hskip 10pt [\hat v_\alpha ^{-}, \hat v_\beta^{-}]=0\eqno( 4.19)$$ Notice that another set of dynamical variables can be constructed from the zeros of $B(\lambda)$ at $ \lambda=-\hat u_\alpha,  \alpha=1,2...N$ and obeying 
$$[\hat w_\alpha ^{\pm }, \hat u_\beta ]=\pm \eta \hat w_\alpha ^{\pm }\delta _{\alpha \beta } \hskip 10pt [\hat w_{\alpha }^{+}, \hat w_\beta^{+}]=0 \hskip 10pt [\hat w_\alpha ^{-}, \hat w_\beta^{-}]=0\eqno( 4.20)$$  where 
$\hat w_\alpha ^{+}=A(\lambda\Rightarrow -\hat u_\alpha )$ and $\hat w_\alpha ^{-}=D^{*}(\lambda \Rightarrow -\hat u_\alpha )$. Then by using Lagrange interpolation one can rewrite the operators $ A(\lambda)$ and $D^{*}(\lambda)$ in the following manner .
$$A(\lambda )=\sum_{\alpha =1}^{2N} \prod_{\beta =1, \beta \ne \alpha }^{2N} \frac{\lambda -\hat \lambda _\beta }{\hat \lambda _\alpha -\hat \lambda _\beta }\hat \mu_\alpha ^{+} +(-)^N r_N \prod_{\alpha =1}^{2N}(\lambda -\hat \lambda _\alpha )\eqno( 4.21)$$
$$D^{*}(\lambda )=\sum_{\alpha =1}^{2N} \prod_{\beta =1, \beta \ne \alpha }^{2N} \frac{\lambda -\hat \lambda _\beta }{\hat \lambda _\alpha -\hat \lambda _\beta }\hat \mu_\alpha ^{-} +(-)^N r_N \prod_{\alpha =1}^{2N}(\lambda -\hat \lambda _\alpha )\eqno( 4.22)$$
$$ \hat \lambda_\alpha=\hat u_\alpha ,\hskip 10pt 1\le \alpha \le N$$
$$ \hat\lambda_\alpha =-\hat u_\alpha  \hskip 10pt  N+1\le \alpha \le 2N$$
$$ \hat \mu_\alpha^{\pm }= \hat v_\alpha ^{\mp } \hskip 10pt 1\le \alpha \le N$$
$$ \hat \mu_\alpha^{\pm }= -\hat w_\alpha^{\pm }  \hskip 10pt   N+1\le \alpha \le 2N$$ The transfer matrix then assumes the form:
$$ 2\hat u_\alpha \tau (\hat u_\alpha )=\xi _{+} (2\hat u_\alpha +\eta ) v_\alpha ^{-} +\xi _{+} v_\alpha ^{+}\eqno( 4.23)$$ Now by extensively using the preperties of quantum determinants [14, 9] it can be shown that from the commutation relations for the operators $ \hat u_\alpha , v_\alpha ^{\pm }, w_\alpha ^\pm $ one can define a Hilbert space where the eigenvalue problem reduces to 
$$2u_\alpha \tau (u_\alpha )\Psi ( u_1,.....u_N) =\xi_{+}(2u_\alpha -\eta)\Delta^{-}(u_\alpha )\Psi(....., u_\alpha  -\eta,....) + \xi_{+}\Delta^{+}(u_\alpha)\Psi(..., u_\alpha +\eta,....)\eqno( 4.24)$$ where 
$$\Delta^{-}(u)=\xi_{-}+(u-\frac{\eta}{2}) \hskip 30pt \Delta^{+}(u)=(2u -\eta)(\xi_{-}-(u+\frac{\eta}{2}))\eqno( 4.25)$$ Setting 
$$\Psi (u_1,...u_N)=\prod_{\alpha=1}^N\phi (u_\alpha )\eqno( 4.26)$$ we have finally
$$2u_\alpha \tau (u_\alpha )\phi (u_\alpha )=\xi_{+}(2u_\alpha +\eta)\Delta^{-}(u_\alpha )\phi (u_\alpha -\eta ) +\xi_{+}\Delta^{+}\phi (u_\alpha +\eta )\eqno(4.27)$$
\section{Relation with Baxter's Q-operator}
In this section we consider a quantum mechanical version of the B\"{a}cklund transformation and show how in the quasi periodic case it may be connected to Baxter's Q-Operator. For this purpose we quantise the dynamical variables occurring in the Lax operator given in (1.3) which then asssumes the form.
$$ L_i(\lambda )=\left(\begin{array}{cccc}
\lambda -\eta q_i \partial _{q_i} & q_i\\ -\eta \partial _{q_i} & 1\end{array}\right)\eqno( 5.1)$$ where we have set $ r_i=-\eta \partial _{q_i}$ and $ [q_i, r_j]=\eta \delta _{ij}$. The quantum analog of the BT for integrable DST model which essentially has the same Lax operator as ours was studied in great detail by Kuznetsov, {\it et al} in [15]; others  have also studied the same model from alternative points of view [16, 11]. The essential idea is to reproduce an operator satisfying the following conditions,
$$ \hat  J(\lambda )\hat Q(\lambda )= \Delta _{+}(\lambda )\hat Q(\lambda +\eta )+\Delta_{-}(\lambda )\hat Q(\lambda -\eta )\eqno(5.2)$$ 
$$\hat Q(\lambda )\hat Q(\mu )=\hat Q(\mu )\hat Q(\lambda )\eqno(5.3)$$
$$\hat J(\mu )\hat Q(\lambda )=\hat Q(\lambda )\hat J(\mu )\eqno( 5.4)$$ and having a common set of eigenfunctions $$ \hat Q(\lambda )\Phi =Q(\lambda )\Phi , \hskip 30pt \hat J(\lambda )\Phi =J(\lambda )\Phi \eqno(5.5)$$ with the eigenvalues satisfying a finite difference equation.
$$ \tau (\lambda )Q(\lambda )=\Delta_{+}(\lambda )Q(\lambda +\eta) + \delta_{-}(\lambda )Q(\lambda -\eta)\eqno( 5.6)$$ Incidentally the last equation arose in Baxter's analysis of the XXX model where $\tau(\lambda)$ was the transfer matrix of the model with $Q(\lambda)$ being a suitable polynomial.\\ To prove (5.3) it is sufficient to demand the existence of an operator $\hat R_{\sigma }$ such that
$${\cal G}(\lambda -\sigma ; t ,\partial _t)L(\lambda; y,\partial_y)\hat R_{\sigma}=\hat R_{\sigma}L(\lambda; q,\partial_q){\cal G}(\lambda-\sigma; s, \partial_s)\eqno(5.7)$$ where we have dropped the indices $'i'$ on the variables for brevity. Note that here ${\cal G}(\lambda-\sigma)$ is assumed to satisfy the following algebra.
$$R_{12}(\lambda -\mu ){\cal G}^1 (\lambda ){\cal G}^2(\mu )=){\cal G}^2(\mu )){\cal G}^1 (\lambda )R_{12}(\lambda -\mu )\eqno(5.8)$$ and as before $$R(\lambda-\mu)={\cal I} +\frac{\eta}{\lambda-\mu}{\cal P}\eqno( 5.9)$$ regarding the explicit form of ${\cal G}(\lambda-\sigma)$ we take it to be as follows.
$${\cal G}(\lambda-\sigma; s,\partial_s)=\left(\begin{array}{cccc}
\lambda- \sigma -s\partial_s & s\\ -\eta\partial_s & 1\end{array}\right)\eqno( 5.10)$$ On the basis of the above remarks proceeding as in [15] we finally arrive at a form of $\hat Q(\sigma)$ as an integral operator with a kernel given by$$ {\cal Q}(\sigma)=\prod_{i=1}^N w_i(\sigma; y_{i+1}, y_i, q_i)\eqno(5.11)$$ where $$w_i(\sigma; y_{i+1}, y_i, q_i)=\frac{i}{2\pi}\Gamma(\sigma/\eta +1) y_i^{-1}(\frac{y_{i+1}-q_i}{y_i})^{-\frac{\sigma}{\eta}-1}exp(\frac{y_{i+1}-q_i}{\eta y_i})\eqno( 5.12)$$ To derive the last property of Baxter's Q-operator we note that $\hat J(\lambda)=tr [C(\lambda, \xi)T_N(\lambda)]$ in the quasi periodic case.  So we consider
$$[\hat Q(\sigma)\hat J(\sigma)\Phi](\vec y)=tr \int dq^N {\cal Q}(\sigma)T_N(\sigma)\Phi(\vec q)\eqno( 5.13)$$ which yields for the kernel of the joint operator $\hat Q(\sigma)\hat J(\sigma)$
$$[\hat Q(\sigma)\hat J(\sigma)\Phi](\vec y\mid \vec x)=tr \{C(\xi)T_N^{*}(\sigma)\prod_{i=1}^N w_i\}= tr \{C(\xi)\prod_{i=1}^N L^{*}(\sigma) w_i\}\eqno(5.14)$$ where $L^{*}(\sigma)$ represents the adjoint of the corresponding local Lax operator. The last equation may be recast into the following form
$$[\hat Q(\sigma)\hat J(\sigma)\Phi](\vec y\mid \vec x)={\cal Q}(\sigma)tr [C(\xi)\tilde T_N(\sigma)]\eqno( 5.15)$$ with $$\tilde T_N(\sigma)=\prod_{i=1}^N\tilde L_i(\sigma)=\prod_{i=1}^N\left(\begin{array}{cccc}
\sigma +\eta +\eta q_i\partial_{q_i}\log w_i & q_i\\ \eta\partial_{q_i}\log w_i & 1\end{array}\right)\eqno( 5.16)$$
It is now possible to triangularise the matrix $\tilde L_i(\sigma)$ by the following transformation:
$$\tilde L_i(\sigma)\longrightarrow S_{i+1}^{-1}\tilde L_i(\sigma)S_i, \hskip 30pt S_i=\left(\begin{array}{cccc}
1 & y_{i+1}\\ 0 & 1\end{array}\right)\eqno(5.17)$$ This transformation leaves the quasiperiodic boundary matrix $C(\xi)$ invariant, while causing $\tilde L_i(\sigma)$ to become triangular as shown below.
$$ \tilde L_i(\sigma)\longrightarrow \left(\begin{array}{cccc}
\sigma \frac{w_i(\frac{\sigma}{\eta} -1)}{w_i(\frac{\sigma}{\eta})} & 0\\
\star & \frac{w_i(\frac{\sigma}{\eta} +1)}{w_i(\frac{\sigma}{\eta})} \end{array}\right)\eqno(5.18)$$  It is then straightforward to show that 
$$ J(\sigma, \xi)\prod_{i=1}^N w_i(\frac{\sigma}{\eta}) =\xi^{-\frac{1}{2}}\sigma^N \prod_{i=1}^N w_i(\sigma/\eta -1) +\xi^{\frac{1}{2}}\prod_{i=1}^N w_i(\sigma/\eta +1)\eqno( 5.19)$$ This relation ought to be compared with the eigenvalue deduced by the standard method of algebraic Bethe ansatz for the quasi periodic case which is of the form:
$$ \Lambda (\sigma)\prod_{i=1}^m (\sigma -\mu_i)=\xi^{-\frac{1}{2}}\sigma^N \prod_{i=1}^m (\sigma -\mu_i -\eta) + \xi^{\frac{1}{2}}\prod_{i=1}^m (\sigma -\mu_i +\eta)\eqno(5.20)$$ 
\section{Discussion}
We have analysed a discrete nonlinear integrable system in the presence of finite boundary conditions and have shown how the diagonalisation problem can be handled when the boundary matrices are upper  (lower) triangular in nature. The construction of B\"{a}cklund transformations under different types of boundary conditions was analysed and for the quasiperiodic case we have shown that a connectionbetween the quantised form of the BT and Baxter's Q operator may be established.\\
One of  us (AGC) wishes to acknowledge the financial support provided by UGC ( Govt. of India)  PSW-026/99-00.
\section{References}
$[1]$ E.K. Sklyanin, (1988){\it J. Phys. A. Math. Gen.} {\bf 21}, 2375\\
$[2]$ I. Merola, O. Ragnisco and Tu G. Zhang (1994) {\it Inverse Problems}{\bf 10}1315\\
$[3]$ H. Q. Zhou (1996){\it J. Phys. A math. Gen.} {\bf 29} L607\\
$[4]$ H. Zhang, G.Z.Tu, W. Oevel and B. Fuchssteiner, (1991) {\it J. Math. Phys} {\bf 32} 1908.\\
$[5]$ E.K. Sklyanin, (1987) {\it Func. Anal. and Appl.} {\bf 21}(2), 86\\
$[6]$ V. B. Kuznetsov, and E. K. Sklyanin (1998) {\it J. Phys. A  Math. Gen}{\bf 31}, 2241\\
$[7]$ A. N. W. Hone, V. B. Kuznetsov and O. Ragnisco , Solv-int /9904003\\
$[8]$ L.D. Faddeev and L.A. Takhtajan, (1979) {\it Russian math. Sueveys} {\bf 34} ( 5), 11; V. A. Korepin, N. M. Bogoliubov and A. G. Izergin {\it Quantum Inverse Scattering Method and Correlation Functions}, Cambridge University Press, 1993\\
$[9]$ E. K Sklyanin in Nankai lactures on Mathematical Physics " {\it Quantum groups and Quantum Integrable Systems}" World Scientific Ed. Mo-Lin Ge pp 63  (1992)\\
$[10]$ V. B. Kuznetsov, M. F. Jorgensen and P. L.  Christiansen 91995) {\it J. Phys. A Math  Gen.} {\bf 28} 4639\\
$[11]$ V. Pasquier, M. Gaudin (1992){\it J. Phys. A. Math.  Gen} {\bf 25}, 5243; R. J. Baxter, {\it Exactly Solved models in Statistical mechanics} Academic Press, New York/London (1982)\\
$[12]$ V. Z. Enol'skii, M. salerno, N. A. Kostov and A. C. Scott (1991) {\it Physica Scripta}{\bf 43}, 229\\
$[13]$ V. Z. Enol'skii, V. B. Kuznetsov and M. Salerno (1993){\it Physica D} {\bf 68}, 138\\
$[14]$ K. Hikami, 91996){\it J. Phys. Soc. Jpn}{\bf 65}(5), 1213\\
$[15]$  V. B. Kuznetsov, M. Salerno and E.K.Sklyanin 92000){\it J. Phys. A Math. Gen} {\bf 33}, 171 \\
$[16]$ E.K. Sklyanin in "B\"{a}cklund transformation and Baxter's Q operator" Lect. at the seminar: "Integrable systems from Classical to Quantum" ( Univ. de Montreal, July 26- August 6, 1999) and S. E Derkachov solv-int/9902015\\

\end{document}